# Surface phonons of NiO(001) ultrathin films grown pseudomorphically on Ag(001)

K. L. Kostov,[1,2] S. Polzin,[1] F.O. Schumann,[1] and W. Widdra,[1,3]

[1] Institute of Physics, Martin-Luther-Universität Halle-Wittenberg, 06120 Halle, Germany

[2] Institute of General and Inorganic Chemistry, Bulgarian Academy of Sciences, 1113 Sofia, Bulgaria

[3] Max Planck Institute of Microstructure Physics, 06120 Halle, Germany

For a ultrathin NiO(001) film of 4 monolayers (ML) thickness grown on Ag(001), the vibrational properties have been determined by high-resolution electron energy loss spectroscopy (HREELS). For the well-ordered pseudomorphically grown film, nine phonon modes have been identified and their dispersions have been revealed along the $\overline{\Gamma}\,\overline{X}$ high-symmetry direction. The comparison with phonon data for a 25 ML thick NiO(001) film shows that the NiO (001) phonon properties are already fully developed at 4 ML. Significant differences are found for the surface-localized phonon $S_6$ which has an increased dispersion for the ultrathin film. The dipole-active Fuchs-Kliewer phonon-polariton exhibits a narrower lineshape than the mode found for a single-crystal surface, which might hint to a reduced antiferromagnetic coupling in the ultrathin film.

## 1. Introduction

Lattice vibrations or phonons are low-energy fundamental excitations of any material which are important for many finite temperature processes. Especially for insulators for which low-energy electron-hole excitations are absent they are, e.g., the dominating contribution to the heat capacitance. For NiO which is a charge-transfer insulator with an optical band gap of 4 eV and an antiferromagnetic ordering of type II below 523 K, the bulk phonon dispersion has been studied since more than 50 years [1-3]. However, the details of the phonon dispersion of NiO attracted also recently new attention due to a significant spin-phonon coupling [4-6]. The vibrational properties of the surface of NiO(001) and the specific surface phonon dispersion have been addressed only by a few studies by means of Helium atom scattering (HAS) [7, 8] and high-resolution electron energy loss spectroscopy (HREELS) [9, 10]. Whereas the former has an excellent energy resolution, it is restricted to low-energy phonon modes due the limited kinetic energy of the scattering Helium atoms. Here we present a detailed study of the full surface phonon dispersion of an ultrathin NiO(001) film of only 4 monolayers (ML). Although there are several phonon studies for metal and rare gas thin films [11], to the best of our knowledge, the present work represents the first determination of the surface phonon dispersion within an ultrathin oxide film. In the following, the results of a highly ordered NiO(001) film of 4 ML thickness are discussed, which has been grown pseudomorphically on a Ag(001) substrate. They will be compared to the phonon dispersion of a thick and relaxed NiO(001) film [10] that resembles well the surface of a NiO(001) single crystal. With the recent experimental advances to measure directly the surface stress during the growth of a NiO thin film on Ag(001) by Dhaka et al. [12], it will be possible to relate static stress measurements with the response in the lattice dynamics.

## 2. Experimental

The experiments have been performed in a two-chamber UHV system with a base pressure in the order of $10^{-9}$ Pa. In the preparation chamber the sample has been cleaned by cycles of Ar$^+$ sputtering (3μA at 1 keV ion energy) and annealing to 700 K. Subsequently, the oxide layers have been grown and characterized by low-energy electron diffraction (LEED).

In the second chamber, a high-resolution electron energy loss (HREELS) spectrometer (Delta 05, SPECS GmbH, Berlin) is used for the surface-sensitive vibrational spectroscopy. Typical energy resolutions for clean metal surfaces of about 1 meV (~8 cm$^{-1}$) and specular count rate of $10^6$ s$^{-1}$ at 4 eV electron energy have been demonstrated with this setup [13-15]. For the present study, an energy resolution in the range of 2-3 meV (16-25 cm$^{-1}$) is chosen in order to record reliably the very low-intensity phonon features in off-specular measurements. A wide electron-energy range from 4 to 121 eV and off-specular angles up to 48° have been used to access the full first surface Brillouin zone (SBZ) and to reach the $\overline{\Gamma}$ point of the second SBZ. The electron incidence angle is 60° with respect to the surface normal. The scattering plane is adjusted with the help of LEED to be parallel to the [110] surface direction, which corresponds to the $\overline{\Gamma}\,\overline{X}$ direction of the SBZ.

The NiO layer with a thickness of 4 ML has been grown on Ag(001) by Ni evaporation in oxygen pressure of $4\times10^{-7}$ mbar ensuring the complete Ni oxidation for an evaporation rate of 0.2 ML/min at 300 K. The desired thickness of NiO layer has been controlled by the method of reflection high-energy electron diffraction (RHEED) monitoring the (00) diffraction spot intensity during Ni evaporation. Periodic oscillations have been observed indicating a layer-by-layer growth mechanism. Further experimental details have been reported recently in studies of the bare Ag(001) [16] and of thick relaxed NiO layers [10].

## 3. Results and Discussion

### 3.1 Measurements at $\bar{\Gamma}$ point

A HREEL spectrum for 4 ML NiO layer on Ag(001) is shown in Fig. 1 in specular reflection geometry including the energy gain and loss regions. We observe a sequence of phonon peaks both on the energy loss as well as on the energy gain side. The intensity ratio between gain and loss peaks obeys strictly the Boltzmann relation exp(-ℏω/kT), where T is the surface temperature of 300 K and ℏω is the phonon energy. This has been used for a precise deconvolution of the HREEL spectra as shown exemplarily for the different phonon contributions in Fig.1. Here, the most intensive losses at 513.8 and 412.5 cm$^{-1}$ correspond to the excitations of two perpendicularly polarized surface vibrations, the Fuchs-Kliewer (FK) surface phonon polariton [17, 18] and the Wallis phonon mode [19], respectively. The FK phonon intensity shows a strong NiO thickness dependence (not shown here). For 4 ML NiO, it is approximately 10 times lower than the measured intensity for 25 ML NiO on Ag(001) [10].

Besides FK and Wallis modes, a number of weak phonons are also visible under the specular scattering conditions of Fig.1. The existence of some, as e.g. the low-intensity peaks at ~365 and ~150 cm$^{-1}$, are not obvious, but they are clearly distinguishable at off-specular measurements presented later.

### 3.2 Off-specular measurements

Figure 2 presents a series of off-specular spectra with increasing electron momentum transfer in the $\bar{\Gamma}\bar{X}$ high-symmetry direction. The spectra measured with a primary-electron energy of 81 eV show the phonon dispersion in the first SBZ up to the center $\bar{\Gamma}'$ of the second SBZ. The intensity of the dipole-active FK phonon is strongly attenuated in off-specular measurements as expected for this excitation mechanism. On the other hand, the high-energy shoulder (HES) of the FK peak under specular scattering conditions (see also Fig. 1) develops to a separate peak at 567 cm$^{-1}$ and is clearly resolved at $\Delta K_{II}$=0.24 Å$^{-1}$ where its intensity is equal to the main FK loss intensity.

The next interesting feature in Fig. 2 is the relative intense Rayleigh (RW) phonon loss with the lowest vibrational frequency in the spectra and characteristic strong dispersion with maximum frequency at $\bar{X}$ point. Several other phonon losses are also observed. As mentioned above the phonon feature at ~150 cm$^{-1}$ is clearly seen at momentum transfer ΔK$_{II}$=0.48 Å$^{-1}$, especially on the gain side (Fig.2). The phonon mode at ~220 cm$^{-1}$, detected also in Fig. 1 for electron energy of 4 eV, dominates at the boundary region around $\bar{X}$ point of the SBZ. Also, the phonon mode suggested to exist at ~365 cm$^{-1}$ in Fig. 1, now can be resolved and its dispersion is clearly visible at least for momentum transfer values in the range of 0.24-0.84 Å$^{-1}$ (Fig. 2). Such phonon mode was also detected in our previous study on thick NiO layer (25 ML) [10] in good agreement also with the HREEL study of bulk NiO(001) [9]. This vibrational loss has been interpreted as excitation of microscopic optical Lucas mode corresponding to vibration parallel to the surface [20].

As discussed above, the high-energy shoulder of the FK peak at 567 cm$^{-1}$ leads to a separate peak at non-zero momentum transfer. This is clearly seen in Fig. 3a where the HREEL spectra from Fig. 2 at the $\bar{\Gamma}$ and the $\bar{X}$ points are compared. The 567 cm$^{-1}$ peak appears as a single well-separated peak at the boundary of the SBZ allowing a precise determination of its location. This is in good agreement with the deconvolution of the 4 eV spectrum in Fig. 1. The intensities of the main FK loss and the 567 cm$^{-1}$ peak are compared with the specular peak intensity in Fig. 3b as function of the momentum transfer along $\bar{\Gamma}\bar{X}$. The intensity of the elastic peak drops drastically by more than three orders of magnitude with increasing momentum transfer ΔK$_{II}$, which indicates the long-range order of the 4 ML film. A proportional drop of the intensity shows the dipole-active FK. Above 0.6 Å$^{-1}$, the FK is not observed anymore. The 567 cm$^{-1}$ peak intensity drops also up to 0.4 Å$^{-1}$ but slower than the FK peak. In the second half of the SBZ, it even increases slightly as demonstrated in Fig. 3b. This shows that while the FK phonon polariton is excited by the dipole mechanism, the excitation of the 567 cm$^{-1}$ mode occurs substantially via impact scattering.

For thick NiO layers we found a power-law dependence of the normalized FK intensity (I$_{FK}$/I$_0$) versus the electron kinetic energy E$_0$ with I$_{FK}$/I$_0$ ~ E$_0^{-0.61}$[10]. The observed power-law exponent of -0.61 for the thick NiO film is very close to the theoretical expected value of -0.5 for infinite thick layers, whereas an exponent of -1 is expected for a monatomic layer [21]. A similar analysis for the 4 ML HREELS data presented here leads to an E$_0^{-0.74}$ dependence. The exponent of -0.74 signalizes therefore an intermediate behavior between bulk-like NiO(001) and a monolayer oxide.

### 3.3 Surface-phonon dispersions

From the phonon energies at different momentum transfers we construct the full phonon dispersion relation (see below). Additionally, we take advantage of the strong dependence of the inelastic excitation cross sections on electron kinetic energy $E_0$ to resolve close lying phonons [22]. At a given momentum transfer, this leads to large variations of the loss intensities of different phonons with $E_0$. One example is demonstrated in Fig. 4 where HREEL spectra are compared for four different kinetic energies but fixed $\Delta K_{II}$ = 0.25 Å$^{-1}$. Note that all spectra at a given momentum have been fitted with the same set of phonon energies and widths. Only the relative intensities are allowed to vary. Due to the small $\Delta K_{II}$, the dominant losses in Fig. 4 are still the FK and the Wallis mode at 514 and 409 cm$^{-1}$, respectively. Additionally, one can clearly distinguish the Lucas phonon [20] at about 365 cm$^{-1}$. It overlaps partially with the Wallis mode at ~409 cm$^{-1}$. However, since the intensity ratio of Lucas and Wallis modes varies significantly with electron energy a separation of both modes is possible as shown by solid lines in Fig. 4. Note that at the $\overline{\Gamma}$ point the Lucas mode intensity is very weak as can be seen in Fig. 1. Also, the 567 cm$^{-1}$ peak which appears as a shoulder of the FK peak at the $\overline{\Gamma}$ point is now clearly distinguishable as a separate peak (Fig. 4). As we discussed previously [10], the dipole-active FK phonon polariton peak should be not visible for off-specular scattering conditions for a long-ranged ordered oxide surface. Here, the observed FK intensity in off-specular spectra is due to a low-intensity non-momentum conserving scattering from the $\overline{\Gamma}$ point and therefore replicas of the specular $\Delta K_{II}$= 0 Å$^{-1}$ spectra. This proposition is supported by the strong and almost proportional decrease of the intensities of the specular and the FK peaks with scattering angle as shown in Fig. 3b.

Below 350 cm$^{-1}$ three low-frequency phonon features are clearly resolved in Fig. 4: The loss at about 248 cm$^{-1}$ is visible only for $E_0$=81 eV, whereas another two modes at 124 and 185 cm$^{-1}$ are nicely observable on the loss- and gain-sides of the 121 eV spectrum. The fixed constraint between loss and gain peak intensities given by the Boltzmann factor allows identifying the peak at 124 cm$^{-1}$ as shoulders around the specular peak (marked in green in Fig. 4). Similarly, in the top spectrum for 121 eV electron energy a low-frequency peak at 44.5 cm$^{-1}$ (marked in yellow) is recognized. As result, Fig. 4 allows the identification of seven phonon modes at 44.5, 124, 185, 248, 365, 409, and 567 cm$^{-1}$ in addition to the FK mode at 514 cm$^{-1}$.

Figure 5 shows similarly off-specular measurements for a fixed larger momentum transfer of $\Delta K_{II}$=0.61 Å$^{-1}$. Under these conditions, a clear and dominant feature is the Lucas phonon at about 344 cm$^{-1}$ since the near-by Wallis mode has now significantly lower intensity. In addition, the phonon loss at approximately 202 cm$^{-1}$ is clearly resolved. In the low-frequency region below 200 cm$^{-1}$, the intensities on the gain and loss sides in Fig. 5 vary strongly and can only be explained by two additional modes at ~106 and 135 cm$^{-1}$. Actually, the number of observed phonon modes in Fig. 5 is the same like as in Fig. 4, but their frequencies are shifted due to their dispersion. Figures 4 and 5 demonstrate the procedure of phonon identification at only two selected points of the SBZ.

The result of the analysis of all off-specular measurements at 300K and different momentum transfers $\Delta K_{II}$ in $\overline{\Gamma}\,\overline{X}$ direction are collected in Fig. 6 as open circles. In the range from 0 to 2.1 Å$^{-1}$, one can distinguish 9 phonon branches. However, as it has been discussed above, the dispersionless mode at 515 cm$^{-1}$ originates from the replicas of the dipole-active FK phonon polariton due to its relative high-intensity at the $\overline{\Gamma}$ point. Also the 567 cm$^{-1}$ mode does not show any dispersion up to the boundary of SBZ. Of the remaining 7 phonon branches, five phonon branches show similar energies and dispersions as those measured for a thick (25 ML) relaxed NiO(001) film [10]. The latter are marked as solid circle in Fig. 6 and are located at ~180, ~220, ~370, 413 cm$^{-1}$ at the $\overline{\Gamma}$ point, whereas the Rayleigh wave (RW) phonon shows a strong dispersion from about 60 cm$^{-1}$ at 0.35 Å$^{-1}$ to ~172 cm$^{-1}$ at the $\overline{X}$ point. According to this comparison, the lowest frequency branch observed here for the 4 ML NiO(001) is attributed to the Rayleigh phonon, too. It runs parallel to the bottom of the projected bulk acoustic bands showing a strong dispersion with a clear maximum at ~172 cm$^{-1}$ for the $\overline{X}$ point and a subsequent downward dispersion in the second SBZ. This $\overline{X}$ point frequency is almost equal to the RW frequency of 173 cm$^{-1}$ measured for the relaxed thick NiO film [10].

The nearly dispersionless phonon branch at ~180 cm$^{-1}$ ($\overline{\Gamma}$ point) has also a surface resonance character (Fig. 6). It has been also observed for the bulk-like oxide film and follows the edge of the LA phonon band that appears in this frequency region due to the antiferromagnetic order [10].

The fourth phonon mode starting from ~220 cm$^{-1}$ at the $\overline{\Gamma}$ point exhibits a larger dispersion as compared to that for the bulk-like NiO(001) layer and reaches ~285 cm$^{-1}$ at $\overline{X}$ point. It follows the upper edge of the acoustic bulk phonon between 0 and 0.65 Å$^{-1}$ and becomes a true surface phonon above 0.65 Å$^{-1}$. It is assigned here to the S$_6$ surface phonon as proposed previously for bulk-like NiO(001) [8, 10]. Its changing character to a surface localized phonon around the $\overline{X}$ point is clearly demonstrated in Fig. 2 where it develops into a strong and well-defined peak above 0.72 Å$^{-1}$. We note that the deviation in its phonon frequency below 0.65 Å$^{-1}$ for the 4 ML film as compared to bulk-like NiO(001) increases with decreasing momentum. This can be understood in terms of a larger coupling depth for smaller K$_{II}$ in the case of the surface resonance.

In agreement with previous vibrational studies [8-10], the phonon branch dispersing downwards from 370 to 343 cm$^{-1}$ is assigned to the Lucas surface phonon (S$_4$). However, as a surface phonon it should be located in the gap between projected bulk acoustic and optical bands. This does not match with the grey marked projected bulk band dispersion in Fig.6. As we noted earlier [10] this gap is now "filled" by back-folded LA modes due to the antiferromagnetic order. Their low density of states explains why the Lucas mode is visible in the spectra. The phonon branch at 413 cm$^{-1}$ shows a small downwards dispersion from the $\overline{\Gamma}$ point to 398 cm$^{-1}$ at the $\overline{X}$ point (Fig. 6). A similar branch has been observed earlier and is assigned to a localized surface Wallis phonon (S$_2$) in vicinity to the $\overline{\Gamma}$ point, which falls in the gap between projected acoustic and optical bands [8-10]. At higher momentum transfer ΔK$_{II}$, it changes its character to a surface resonance propagating into projected bulk bands. Likewise we observe a strong phonon peak close to the $\overline{\Gamma}$ point and only a weak feature at high ΔK$_{II}$.

In addition to the known phonons of the bulk-like NiO film, the 4 ML data in Fig. 6 exhibit two new phonon branches at 145 and 455 cm$^{-1}$ at the $\overline{\Gamma}$ point. The branch at 145 cm$^{-1}$ around the $\overline{\Gamma}$ point crosses the projected acoustic bulk bands with increasing momentum and therefore should have resonance character. It is visible up to 0.7 Å$^{-1}$ in the SBZ, when it is reaching the RW. The phonon at 455 cm$^{-1}$ that is visible only around the $\overline{\Gamma}$ point is the second mode which has not been observed earlier for bulk-like NiO(001). This branch follows the edge of the projected optical bulk bands and it is visible up to ΔK$_{II}$ =0.4 Å$^{-1}$ with vanishing dispersion (Fig. 6). It has a rather low intensity in off-specular measurements and is observed only under conditions where the FK mode at 514 cm$^{-1}$ is present (see Fig. 1 and Fig. 4). Its nature is not clear yet. We tend to assign it to a small area of the film with 2 ML height instead of 4 ML since the FK phonon polariton frequency is reduced to 455 cm$^{-1}$ for 2 ML NiO (not shown here). Alternatively, it arises from an intrinsically asymmetric lineshape of the FK mode.

So far, the HREELS data have been discussed for room temperature NiO(001). Low-temperature dispersion data recorded at 80 K are summarized in Figure 7. By comparison with the results obtained at 300 K (Figure 6), it is seen that the dispersion curves are very similar for both surface temperatures. For example, within the experimental error the RW frequency at $\overline{X}$ point is found to be ~173 cm$^{-1}$ in nice agreement with the value at 300 K. Differences are observed in the low-frequency region 115-120 cm$^{-1}$ where one dispersionless mode is observed. In addition, weak peaks at ~300 cm$^{-1}$ has been detected. However, these features can be explained with low-frequency vibrations (frustrated modes) of some adsorbate species at 80 K. For example, a physisorbed CO species is detected with C-O stretching frequency at ~2180 cm$^{-1}$ close to that in gas phase. No peak of adsorbed OH groups has been detected in the region up to 4000 cm$^{-1}$ but this cannot completely exclude their presence on the surface.

### 3.4 Phonons at the zone boundaries of the SBZ

The top three spectra in Fig. 8 present a comparison of off-specular HREELS data at the $\overline{X}$ point for bulk-like NiO(001) at 300 K and 4 ML NiO(001) thin films at 80 and 300 K, all measured at an electron energy of 81 eV. The spectra are very similar as might be expected also from the similar phonon dispersions in Figs. 6 and 7. As mentioned in the previous section, the peak of the surface-localized phonon $S_6$ dominates in all three spectra. Between 500 and 600 cm$^{-1}$, we observe for the thick NiO(001) only the replica of the FK at 559 cm$^{-1}$. For the thin film, the FK has been found at a lower frequency of 514 cm$^{-1}$. Therefore, we attribute the peak at 567 cm$^{-1}$ in the second and third spectrum of Fig. 8 to a NiO(001)/Ag(001) interface phonon, which is derived from a bulk LO phonon and which is not detectable for thicker NiO(001) films.

In Fig. 8 an off-specular spectrum recorded at the SBZ boundary at the $\overline{M}$ point is presented in the lowest spectrum for comparison. The spectrum shows less distinct features, but a number of peaks are retrieved by fitting. The RW frequency is visible at low-energy shoulder at 160 cm$^{-1}$. This lower frequency as compared to the $\overline{X}$ point agrees well with the results of the HAS study of Witte et al. for bulk NiO(001) [8]. The strongest contribution to the spectrum under the given condition is found at 193 cm$^{-1}$ close to the Rayleigh mode frequency which is ascribed to the so-called "optical" Rayleigh phonon ($S_1$`) characteristic for bulk-like NiO and also for CoO(001) at the $\overline{M}$ point [8]. The phonon peak at 327 cm$^{-1}$ is clearly visible and may correspond to the Lucas surface phonon ($S_4$) as suggested also in Ref. [8] for a NiO(001) single crystal.

### 3.5 Shape of FK phonon polariton

For a thick NiO film, we have found earlier a broad line shape with a FWHM independent on the incident electron energy [10]. This line shape has been explained by a FK splitting due to the antiferromagnetic ordering of NiO. For the ultrathin NiO film of 4 ML reported here, the shape of the FK loss is definitely narrower but it has a small additional peak or asymmetry at the high-frequency side, as has been also discussed above. In Fig. 9a a comparison of the FK peaks for the ultrathin (4ML) and the thick NiO(001) films is shown under otherwise similar conditions. The spectra are slightly shifted in frequency and are normalized by their FK intensities for better visualization. The comparison clearly shows the narrower FK peak for 4 ML and the presence of the high-energy shoulder due to an additional loss feature at 567 cm$^{-1}$. Cooling the 4 ML NiO film down to 80 K causes a significant narrowing of the shoulder. If analyzed as separate peak as shown in Fig. 9b, its FWHM is reduced to 56 cm$^{-1}$ from an initial width of ~100 cm$^{-1}$ at 300 K. A red shift of approximately 5-6 cm$^{-1}$ can be suggested for this feature at 80 K. Note that at the same time the main FK frequency is also shifted by about 2-3 cm$^{-1}$ in opposite direction, but which is close to the experimental error. The origin of the high-energy shoulder of the FK mode at 567 cm$^{-1}$ peak is a still an open issue: It is located at the upper edge of the LO phonon band of NiO. A split-off FK component due to a spin-phonon interaction as is observed for the NiO bulk [4-6] would suggest a dipole-active mode, which might explain the feature for $\Delta K_{II} < 0.3$ Å$^{-1}$. Its strong broadening with temperature increase from 80 to 300 K could then be related to a loss of antiferromagnetic order at 300 K for the 4 ML NiO film. In fact, the Néel temperature is expected to be reduced to values below 300 K for a reduced NiO thickness of only 4 ML. However, this proposition would require a significant excitation of this mode by impact scattering to explain the clear observation of the 567 cm$^{-1}$ peak at the zone boundary, which is not expected. Note that we rule out adsorbates as origin of the 567 cm$^{-1}$ peak, since high-temperature annealing does not influence the spectra and since it has been observed for many different preparations. Additionally, its width exceeds significantly the expected width of localized adsorbate modes.

### 4. Conclusion

The surface phonon dispersion relations for a 4 ML ultrathin NiO(001) film grown pseudomorphically on Ag(001) is reported. Based on a large set of HREEL spectra recorded at different scattering conditions with respect to the scattering angle and energy, eight surface phonons and surface resonances have been identified. Their dispersions have been determined along the $\bar{\Gamma}\bar{X}$ high-symmetry direction and are compared to the corresponding phonons of a 25 ML thick film on the same substrate. The results demonstrate that already at 4 ML the phonon dispersion is similar to bulk-like NiO(001). Pronounced differences are found for the $S_6$ surface phonon, which shows a stronger dispersion for the thinner and compressed film. A previously not observed high-frequency mode at the upper edge of the LO bulk phonon band has been found that is most pronounced at the $\bar{X}$ zone boundary.

**Acknowledgement**

This work is dedicated to Dietrich Menzel who taught us (W.W. and K.K.) an accurate and methodically as well as conceptual broad approach to surface science.

The support by the German joint research network Sonderforschungsbereich 762 "Functionality of oxidic interfaces" of the Deutsche Forschungsgemeinschaft is gratefully acknowledged.

**Figure captions:**

Fig. 1: High-resolution electron energy loss spectrum under specular electron scattering conditions for 4 ML thin pseudomorphic NiO(001) film on Ag(001) measured with an incident electron energy of 4 eV at 300 K. The contribution of the different phonon modes and their sum are shown with blue and red curves, respectively. Note that all phonons that are visible as energy loss are also present on the energy gain side.

Fig. 2: Off-specular HREEL spectra for a 4 ML NiO(001) film on Ag(001) measured with an electron energy of 81 eV along the $\overline{\Gamma}\,\overline{X}$ direction. The momentum transfer $\Delta k_\parallel$ values are indicated on the left side of the spectra. The HREEL spectra recorded at the high-symmetry points of the SBZ are colored in red.

Fig. 3: (a) Comparison of the loss spectra for a 4 ML NiO(001) film on Ag(001) measured at the $\overline{\Gamma}$ (open circles) and the $\overline{X}$ (solid line) point of the SBZ (electron energy 81 eV); (b) Comparison of the intensities of the elastically scattered electrons (full black circles), the Fuchs-Kliewer phonon-polariton peak (full red circles) and the FK high-energy shoulder (connected open circles) measured at different momentum transfer $\Delta k_\parallel$ along the $\overline{\Gamma}\,\overline{X}$ direction.

Fig. 4: Off-specular HREEL spectra at constant momentum transfer of $\Delta k_\parallel = 0.25$ Å$^{-1}$ but for different electron energies (indicated left). The contribution of the different phonon modes and their sum are shown with blue and red curves, respectively. The lowest-frequency resonance peaks are colored in green, whereas the RW phonon (colored in yellow) is visible only in the top spectrum measured with electron energy of 121 eV.

Fig. 5: Off-specular HREEL spectra at constant momentum transfer of $\Delta k_\parallel = 0.61$ Å$^{-1}$ but for different electron energies (indicated left). The contribution of the different phonon modes and their sum are shown with blue and red curves, respectively.

Fig. 6: Experimental phonon dispersion (open symbols) for 4 ML NiO(001) on Ag(001) along $\overline{\Gamma}\,\overline{X}$ direction (up to the center $\overline{\Gamma}\,'$ of the second SBZ) at 300 K. Solid circles denote previous HREELS results for a relaxed 25 ML NiO(001) film [10]. The solid and open circles of a given color indicate one phonon branch. The hatched areas mark the surface–projected bulk phonon bands as derived from a DFT+U frozen phonon calculations [10].

Fig. 7: Experimental phonon dispersion (open squares) as in Fig. 6 but at a sample temperature of 80 K.

Fig. 8: Off-specular HREEL spectra of a 4 ML NiO(001) film on Ag(001) at the $\overline{M}$ and $\overline{X}$ points measured with electron energy of 81 eV at 300 K (first and third spectra from the bottom). For comparison, the spectra at 80 K and for a thick 25 ML NiO(001) film at 300 K are also shown (the second and the top spectra, respectively).

Fig. 9: (a) Comparison of the dipolar HREEL spectra in the region of the FK mode for 4 ML(blue line) and 25 ML (black open circles) NiO(001) films on Ag(001) measured at 300 K. (b) Dipolar HREEL spectra in the region of the FK mode for 4 ML at 300 K (blue line) and at 80 K (red line). The areas of the FK high-energy shoulder at 300 K and 80 K are filled in blue and red, respectively.

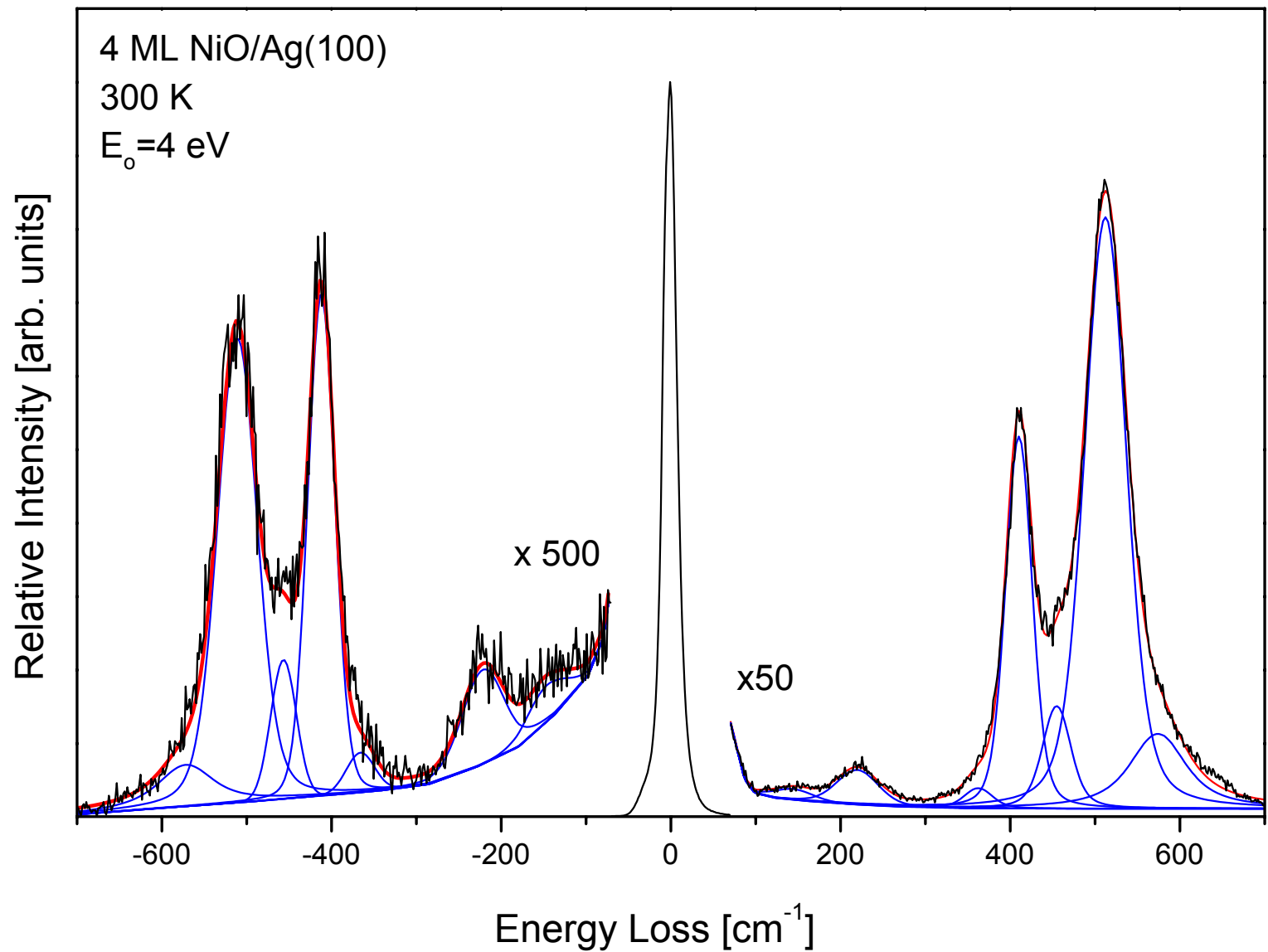

Fig.1

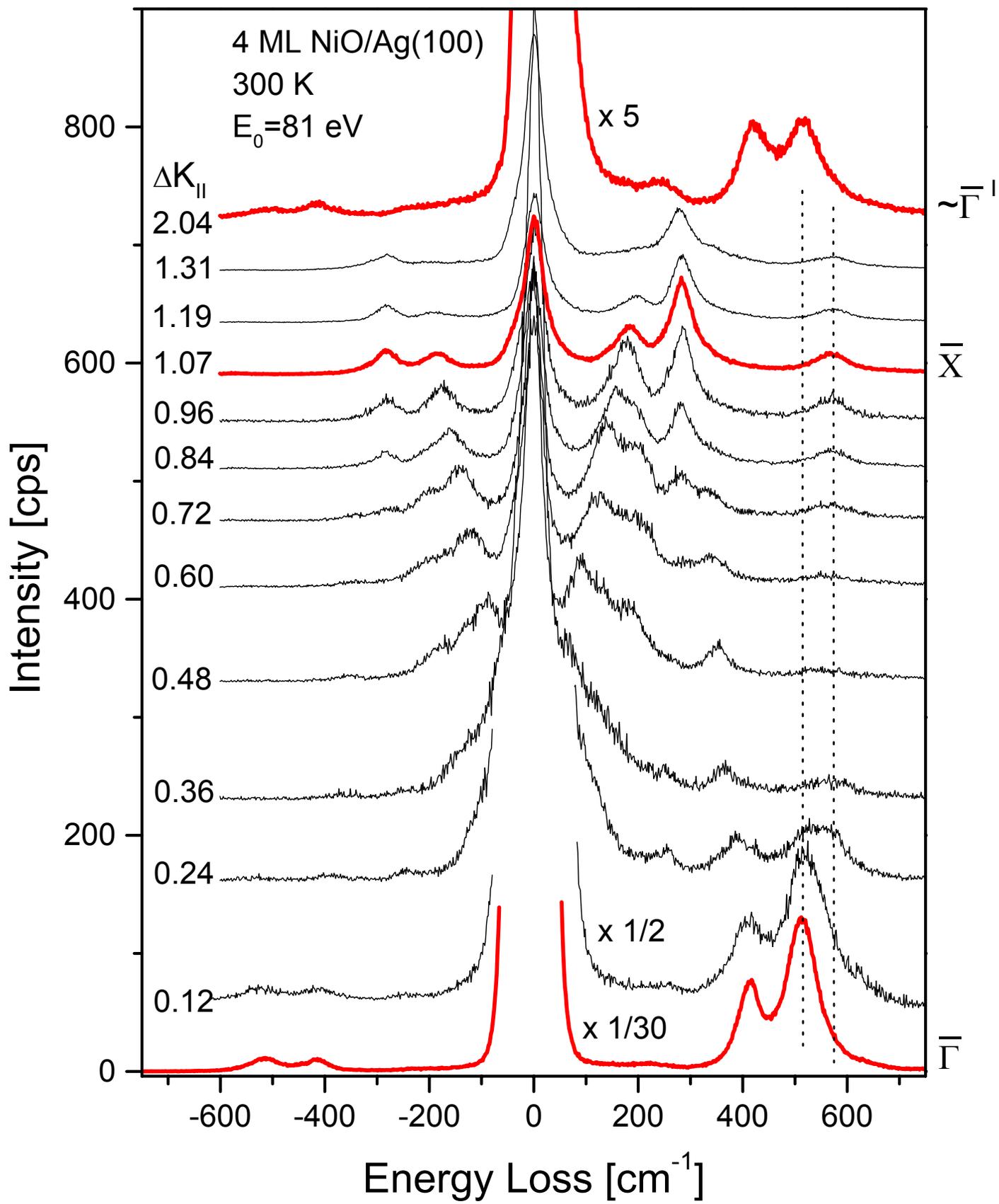

Fig.2

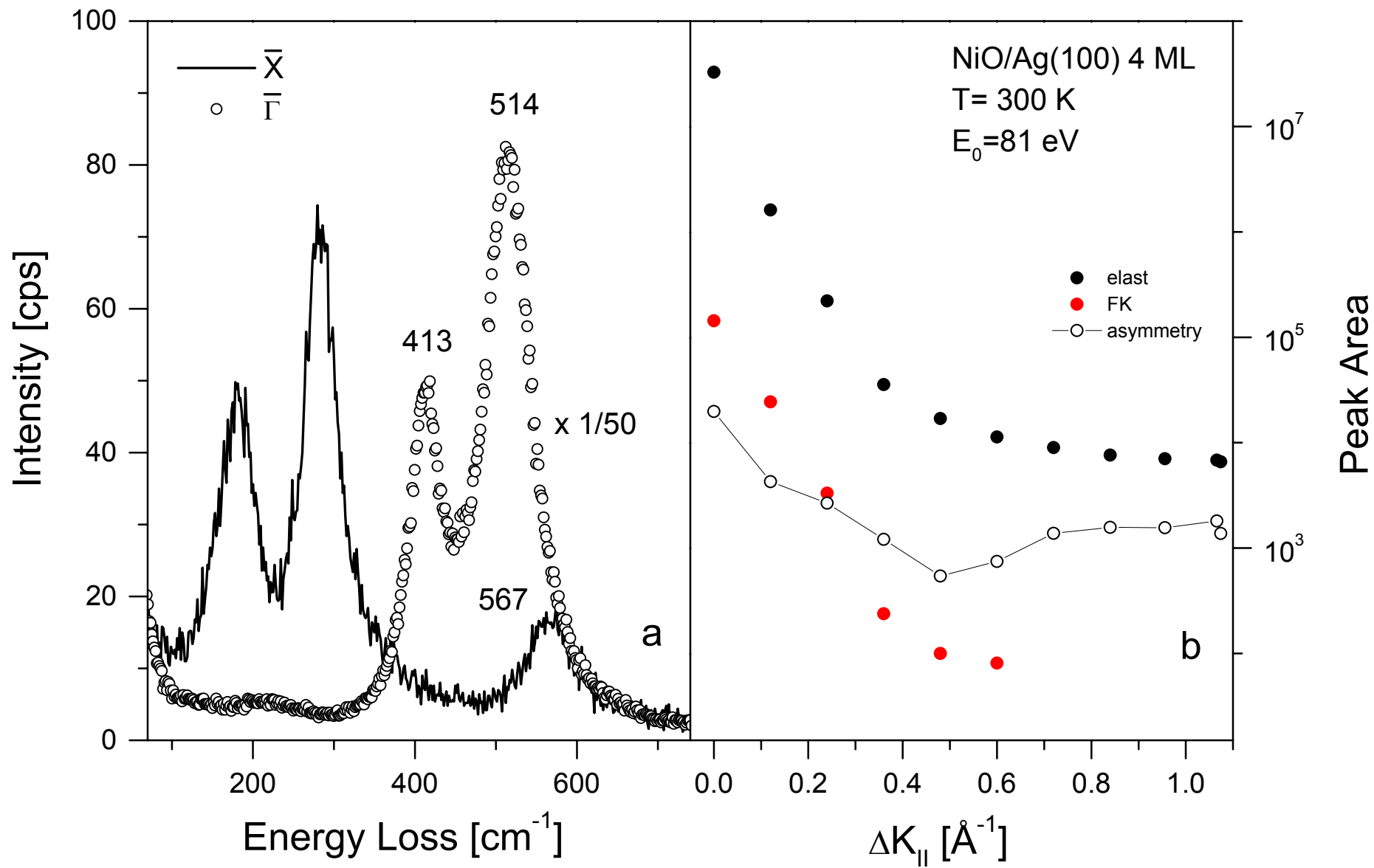

Fig.3

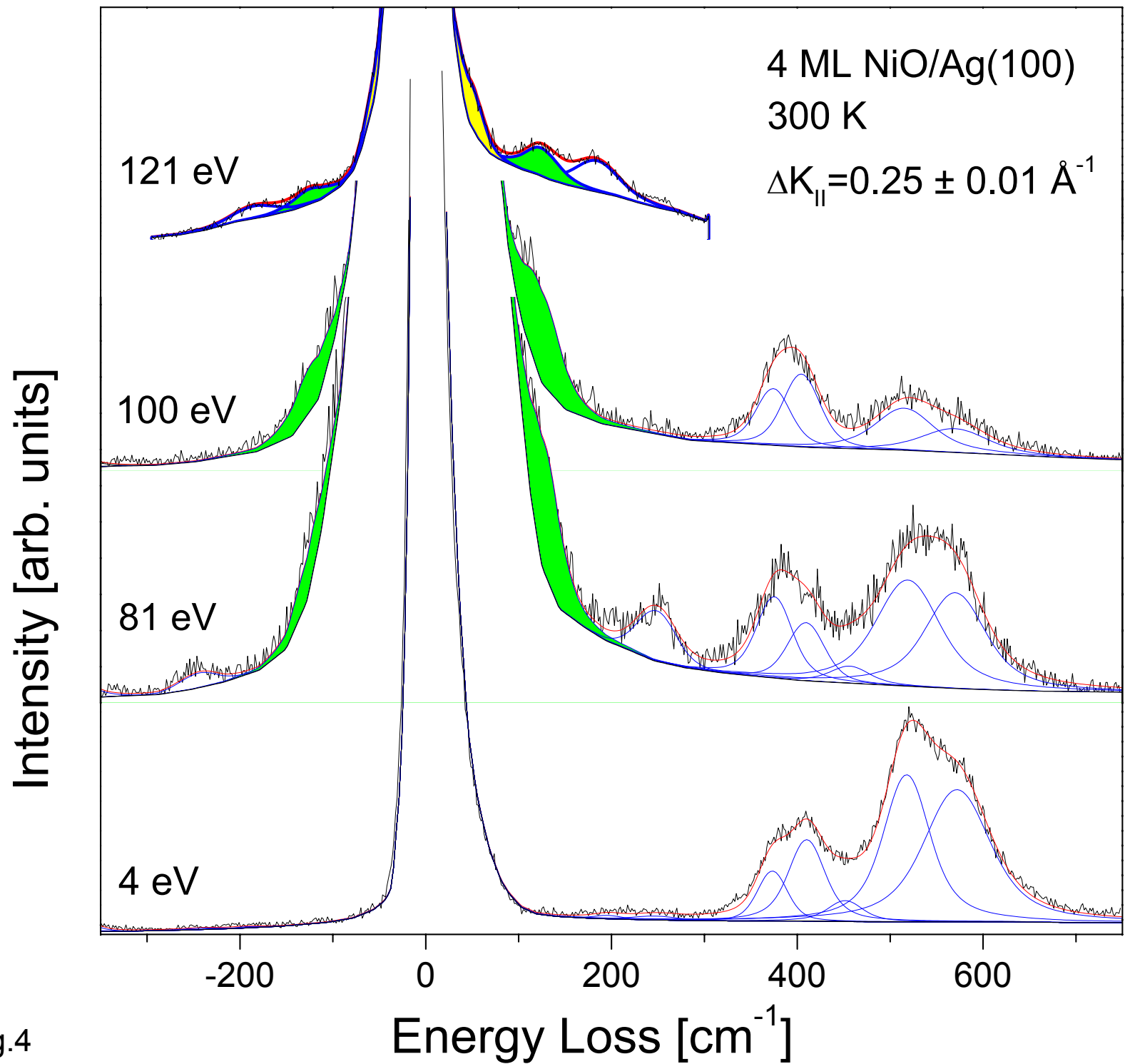

Fig.4

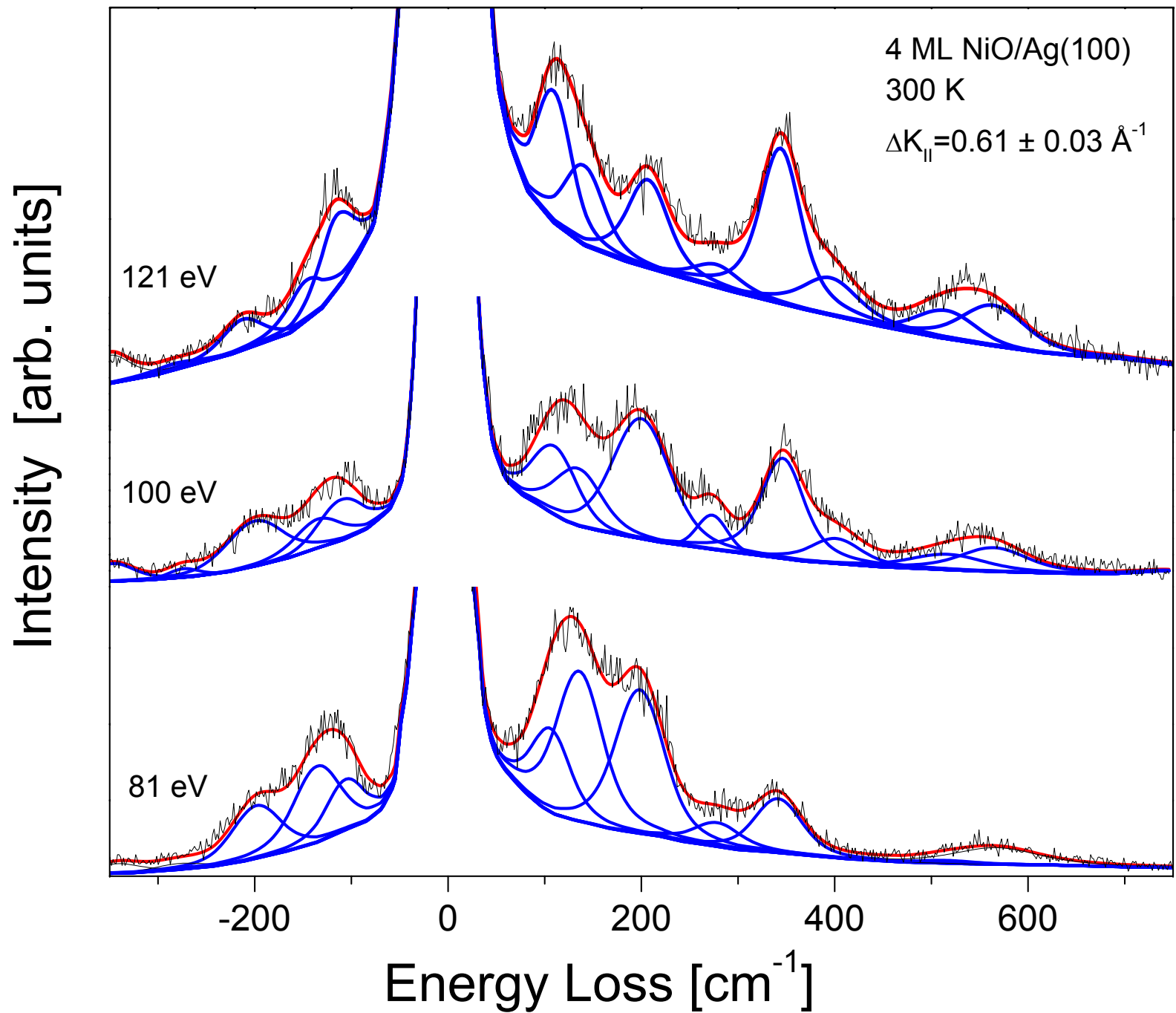

Fig5

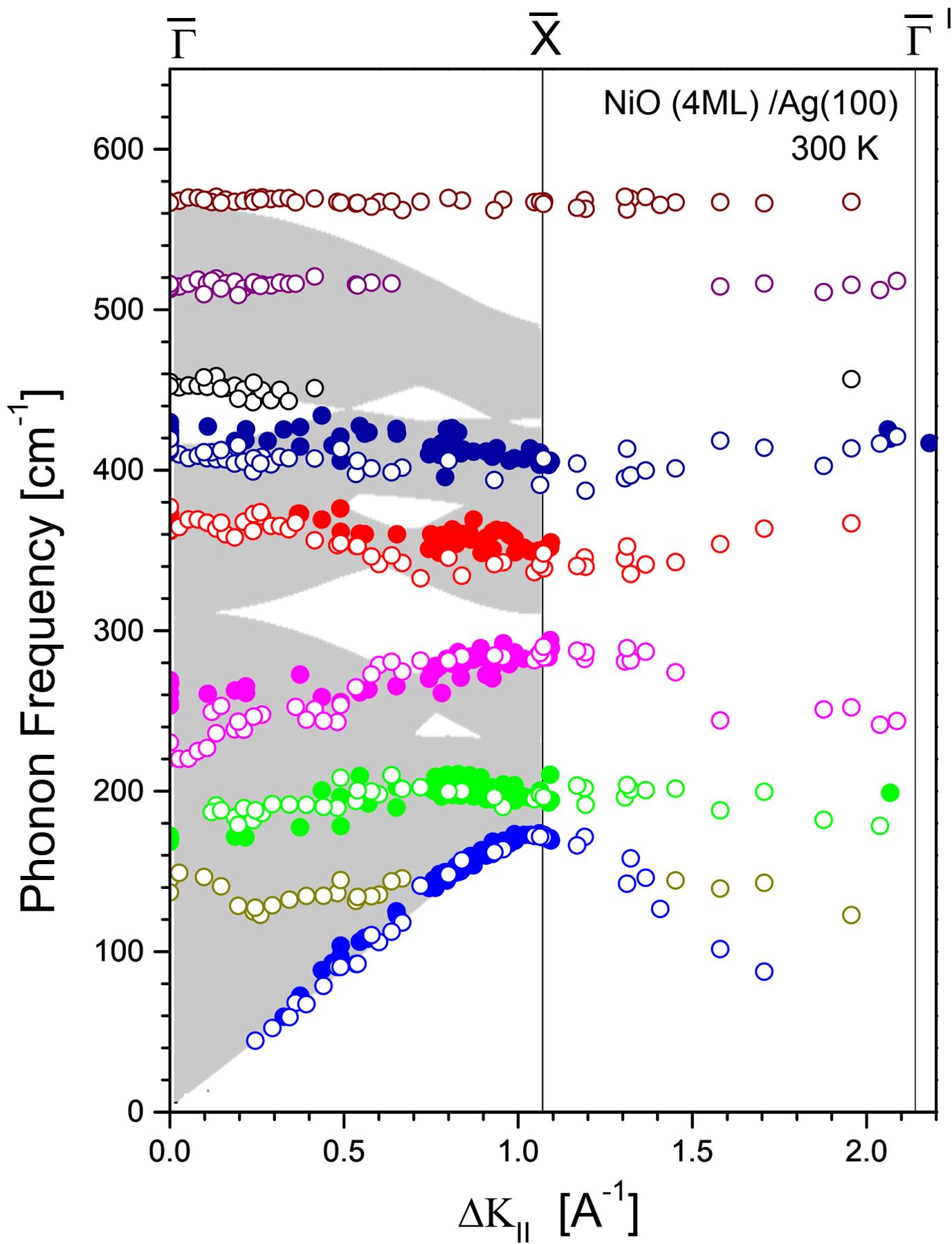

Fig.6

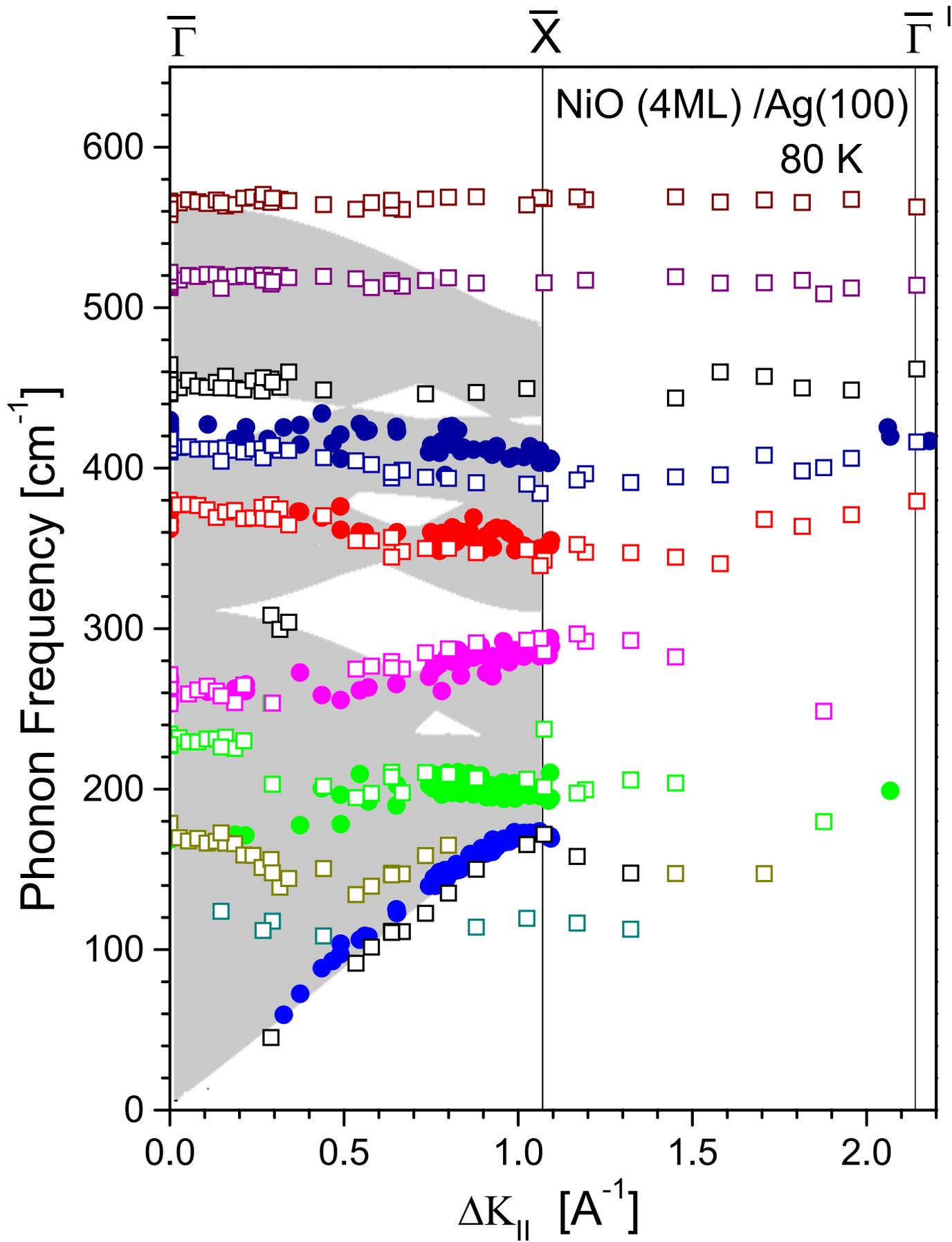

Fig.7

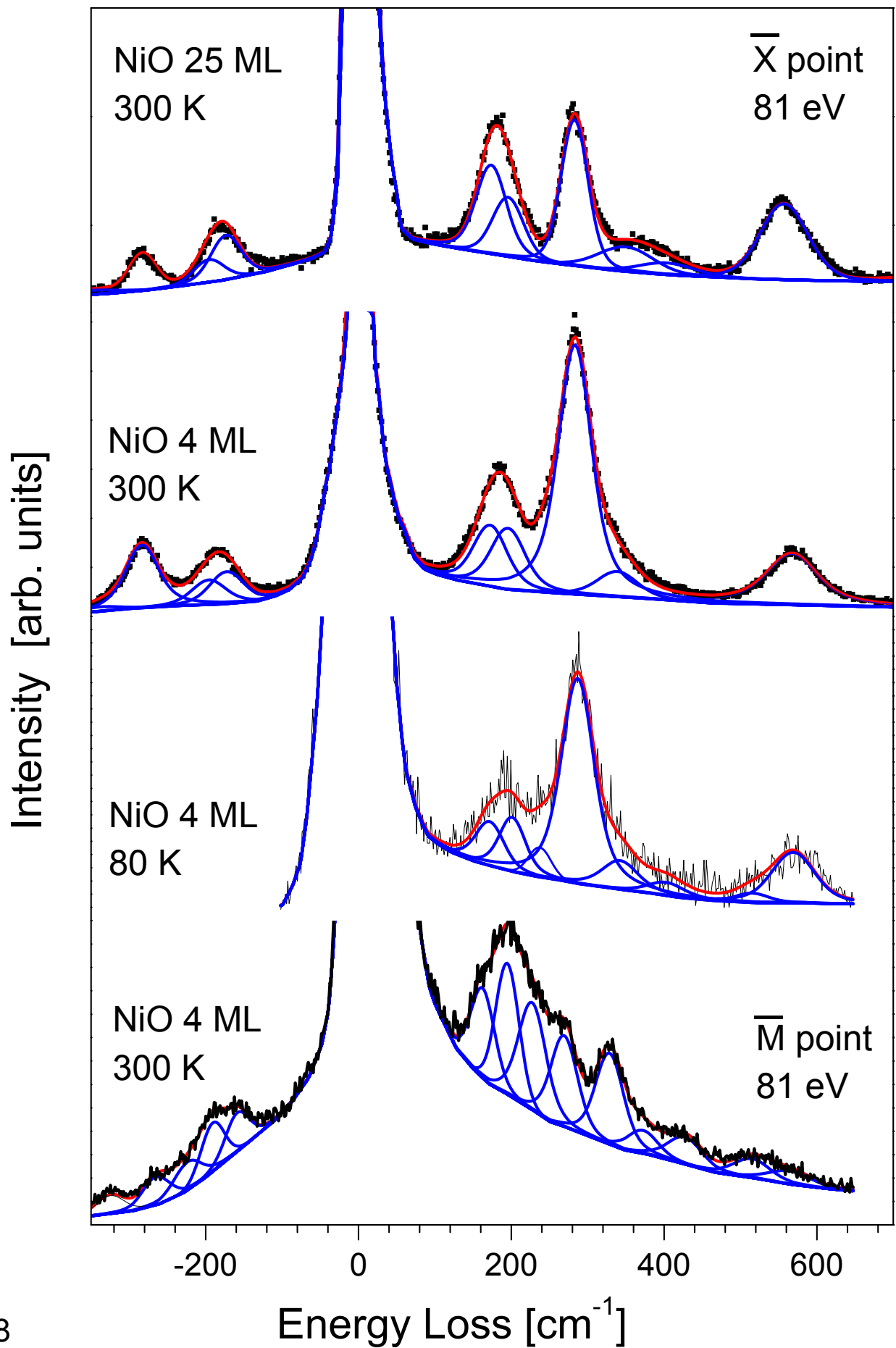

Fig.8

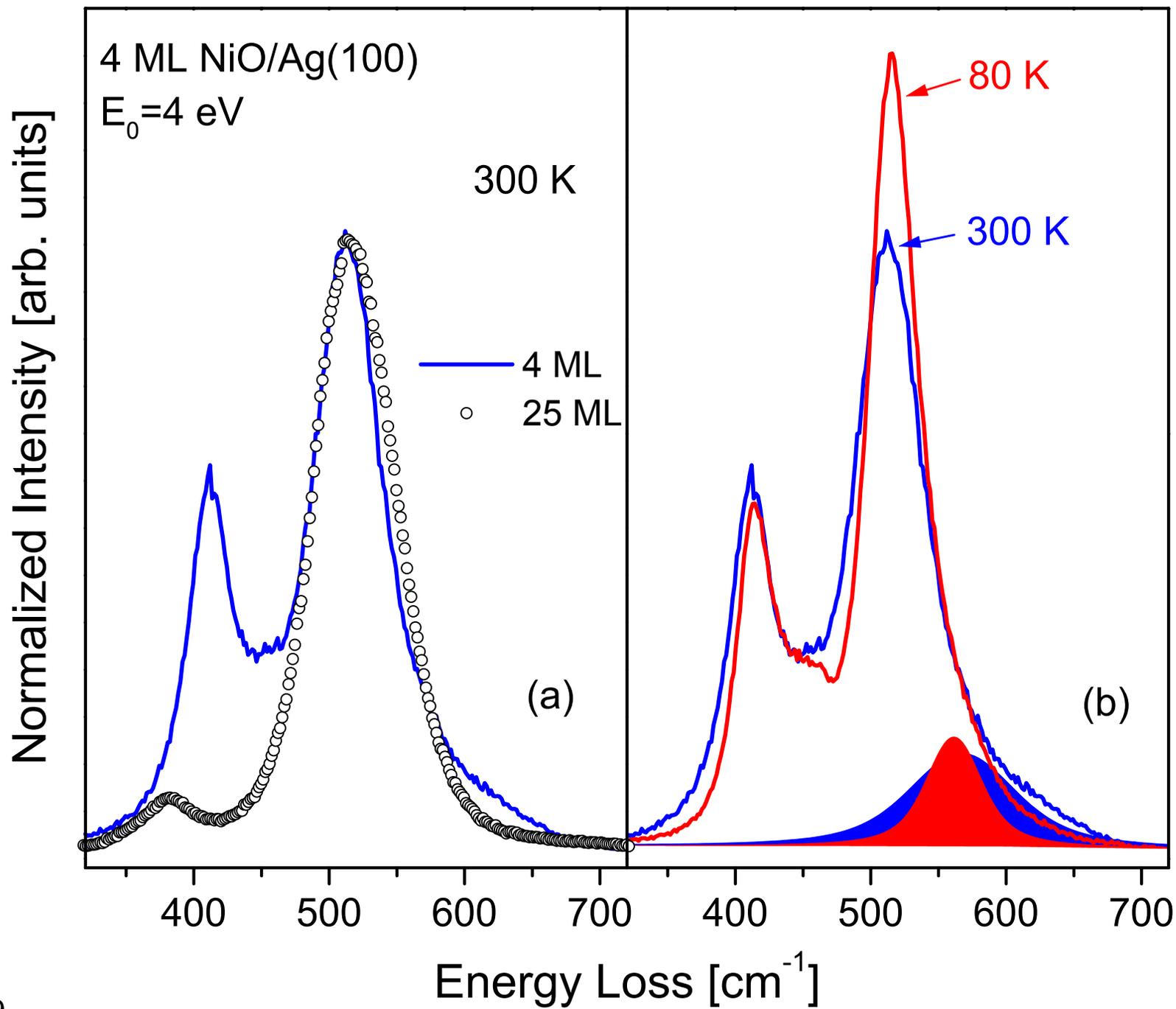

Fig.9